\documentclass[11pt]{article}
\usepackage{axodraw}
\usepackage{epsfig}
\usepackage{amsfonts}
\usepackage{amsmath}
\usepackage{bbm}
\usepackage{bm}
\input pix.sty

\renewcommand{\eq}{Eq.~}
\renewcommand{\eqs}{Eqs.~}
\def\Eq#1{Eq.~(\ref{#1})}
\renewcommand{\se}{Sec.~}

\renewcommand{\fig}{Fig.~}

\newcommand{\tinymsbar}{{\overline{\mbox{\tiny\rm{MS}}}}}
\newcommand{\Lambdamsbar}{{\Lambda_\tinymsbar}}
\newcommand{\betaL}{\beta_{\rmi{L}}}
\newcommand{\betaLl}{\beta_{\rm{L}}}
\newcommand{\mD}{m_\rmi{D}}

\newcommand{\Sigmavi}{\Sigma_{v^{-1}}}
\newcommand{\Cf}{C_{\rm F}}
\newcommand{\Nf}{N_{\rm f}}
\newcommand{\Nc}{N_{\rm c}}

\newcommand{\rmO}{{\mathcal{O}}}

\def\lsi{\raise0.3ex\hbox{$<$\kern-0.75em\raise-1.1ex\hbox{$\sim$}}}
\def\gsi{\raise0.3ex\hbox{$>$\kern-0.75em\raise-1.1ex\hbox{$\sim$}}}
\newcommand{\lsim}{\;\lsi\;} 
\newcommand{\gsim}{\;\gsi\;} 

\newcommand{\nB}{n_\rmii{B}}

 \renewcommand{\nB}[1]{n_\rmii{B{#1}}}
\newcommand{\rmii}[1]{{\mbox{\tiny\rm{#1}}}}

\newcommand{\im}{\mathop{\mbox{Im}}}

\newcommand{\Tint}[1]{{\hbox{$\sum$}\!\!\!\!\!\!\!\int\,}_{\!\!\!\!\raise-0.9ex\hbox{$\scriptstyle{#1}$}}}
\newcommand{\Tinti}[1]{{{\Sigma}\!\!\!\!\raise0.3ex\hbox{$\int$}_\rmii{${#1}$}}}

\newcommand{\bi}{\begin{itemize}}
\newcommand{\ei}{\end{itemize}}

\newcommand{\hide}[1]{ }

\def\ring{\mathaccent"7017} 

\makeatletter \@addtoreset{equation}{section} \makeatother
\renewcommand{\theequation}{\arabic{section}.\arabic{equation}}
\makeatletter
\renewcommand\section{\@startsection {section}{1}{\z@}%
                                   {-5.5ex \@plus -1ex \@minus -.2ex}
                                   {2.3ex \@plus.2ex}%
                                   {\normalfont\large\bfseries}}
\renewcommand\subsection{\@startsection{subsection}{2}{\z@}%
                                     {-3.25ex\@plus -1ex \@minus -.2ex}%
                                     {1.5ex \@plus .2ex}%
                                     {\normalfont\normalsize\bfseries}}
\renewcommand\thesection {\@arabic\c@section}
\renewcommand\thesubsection   {\thesection.\@arabic\c@subsection}
\renewcommand{\@seccntformat}[1]{%
\csname the#1\endcsname.\hspace{1.0em}}
\makeatother

\begin{document}

\begin{titlepage}
\begin{flushright}
BI-TP 2009/03\\
MS-TP-09-1\\
\vspace*{1cm}
\end{flushright}
\begin{centering}
\vfill

{\Large{\bf
Heavy Quark Thermalization in Classical Lattice Gauge  \\[1.5mm] 
Theory: Lessons for Strongly-Coupled QCD
}}

\vspace{0.8cm}

Mikko~Laine$^\rmi{a}$, 
Guy D.~Moore$^\rmi{a,b}$, 
Owe Philipsen$^\rmi{c}$,  
Marcus Tassler$^\rmi{c}$ 

\vspace{0.8cm}

$^\rmi{a}$%
{\em
Faculty of Physics, University of Bielefeld, 
D-33501 Bielefeld, Germany\\  }

\vspace*{0.3cm}

$^\rmi{b}$%
{\em
Department of Physics, McGill University, 
Montr\'eal, QC H3A 2T8, Canada\\  }

\vspace*{0.3cm}

$^\rmi{c}$%
{\em
Institute for Theoretical Physics, University of M\"unster, 
D-48149 M\"unster, Germany \\ }

\vspace*{0.8cm}

\mbox{\bf Abstract}
 
\end{centering}

\vspace*{0.3cm}
 
\noindent
Thermalization of a heavy quark near rest is 
controlled by the correlator of two electric fields along 
a temporal Wilson line. We address this correlator within 
real-time, classical lattice Yang-Mills theory, and elaborate 
on the analogies that exist with the dynamics of hot QCD. In the 
weak-coupling limit, it can be shown analytically that the dynamics 
on the two sides are closely related to each other. For intermediate
couplings, we carry out non-perturbative simulations within the 
classical theory, showing that the leading term in the weak-coupling 
expansion  significantly {\em underestimates} the heavy quark thermalization 
rate. Our analytic and numerical results also yield a general
understanding concerning the overall shape of the spectral function 
corresponding to the electric field correlator, which may be 
helpful in subsequent efforts to reconstruct it from Euclidean 
lattice Monte Carlo simulations. 

\vfill


\vspace*{1cm}
  
\noindent
April 2009

\vfill

\end{titlepage}

%
\section{Introduction}
\la{se:intro}

The real-time dynamics of heavy ion collisions 
is governed by QCD at relatively large coupling, which 
remains poorly understood despite significant theoretical efforts.
While in principle both the weak-coupling expansion and lattice QCD
provide systematically improvable schemes for the calculation 
of any physical quantity, including
unequal-time correlation functions ($\Delta t \gg 1/T$)
at a finite temperature $T$ ($T \gsim 200$~MeV), both are 
in practice faced with serious limitations, related to the 
convergence of the weak-coupling expansion and to the 
need to carry out analytic continuation, respectively.
Therefore many current attempts to describe the real-time
dynamics of QCD at realistic temperatures rely on models. 
For instance, for heavy quark thermalization and diffusion, the main
topics of the present paper, a relatively successful model treatment can be 
obtained by incorporating bound states as dynamical degrees of 
freedom into the description of an otherwise partonic
medium~\cite{rapp}. Unfortunately, such models need typically 
to be tuned to the particular observable in question, rather than 
having a universal character, and also do not allow for 
a systematic improvement.

As an alternative to models,
much recent literature has focused on {\em analogue theories}, 
by which we mean well-defined frameworks which are sufficiently close
to QCD that most interesting QCD measurables have equivalents in the
analogue, but which are nevertheless more amenable to calculation.
Two such frameworks have been especially widely used:
QCD truncated to the first non-trivial order in the weak-coupling 
expansion (see, e.g.,\ ref.~\cite{mt}), 
and ${\cal N}{=}4$ Super-Yang-Mills theory in the limit 
of an infinite number of colors and a large 't Hooft coupling
(see, e.g.,\ refs.~\cite{sea,ct}).

QCD truncated to the first non-trivial order in the weak-coupling
series is the starting point of a systematic expansion, and thus
arguably the most similar analogue theory, guaranteed to be correct
in the limit of a high temperature. The problem is that in the present
setting it is technically extremely hard to work out subsequent terms in 
the weak-coupling series, given that extensive resummations are 
needed for dynamical quantities evolving over long time scales, and 
that in general several terms in the expansion would be needed, 
in order to obtain any kind of convergence. In fact, 
even though five subleading orders are available for 
thermodynamic (i.e.\ equal-time) quantities such as 
the pressure~\cite{QCDpressure}, convergence remains debatable~\cite{lattg7}; 
in the dynamical case at most the first non-trivial order has been 
reached so far~\cite{plasmon,CaronMoore,fermion_mass,Caron_qhat}, 
and the results certainly display very large ${\cal O}(g)$ corrections.  
This appears to indicate that the weak-coupling expansion 
is {\em not} well behaved except at very high temperatures, 
casting doubts on the physical relevance of truncated results 
in the realistic temperature range. 

Super-Yang-Mills theory (SYM) resembles QCD in that it is a gauge theory 
with matter.  However it contains many more matter multiplets than 
ordinary QCD, and in a different representation of the gauge 
group, which makes the matching between the theories 
ambiguous~\cite{match}.
Furthermore, computations on the SYM side are simple (using the
famous AdS/CFT correspondence~\cite{AdSCFT}) only in the limit 
of an infinite coupling, whereas the interesting regime is probably 
intermediate coupling.  

In this paper we argue in favor of another analogue 
theory for the real-time dynamics of QCD: classical Yang-Mills 
theory regulated on a spatial lattice.  This theory was developed 
by Kogut and Susskind~\cite{KogutSusskind} and has been used 
extensively to study the rate of Chern-Simons number diffusion 
in Yang-Mills theory~\cite{Ambjornetal},
as well as (partly) out-of-equilibrium phenomena such as plasma 
instabilities~\cite{instab}, the dynamics of electroweak symmetry 
breaking~\cite{SmitTang}, and 
inflationary preheating~\cite{Garciabellido}.
Recently, it was also applied for estimating the imaginary 
part of the real-time heavy-quark potential 
in QCD~\cite{LainePhilipsenTassler}, and analogous
methods were used for studying jet energy loss and transverse
broadening in a hot non-Abelian plasma~\cite{fra}.
In this approach the infrared (IR) behavior 
of QCD is approximated by introducing semiclassical fields, 
while in the ultraviolet (UV) the quantum mechanical ``cutoff'' 
on thermal effects from short distances is replaced with a 
lattice cutoff.
Formally, the classical limit corresponds 
to taking $\hbar\to 0$, 
which is a non-trivial limit at 
non-zero temperatures~\cite{hbarsquared,sft}, and non-singular
in the presence of the lattice cutoff.

In this paper we use this framework to study one of the simplest 
gauge-invariant observables, the correlator of two electric fields along a
Wilson line,
\begin{equation}
 \kappa(\omega) \equiv 
 \frac{\fr13 \sum_{i=1}^3 {\displaystyle 
   \int \! {\rm d}t \,  e^{i\omega t}  \;
 \tr \left\langle U(-\infty-i\beta,t) \; gE_i(t,\vec{0}) \; U(t,0) \; 
 gE_i(0,\vec{0}) \; U(0,-\infty) \right\rangle }}
 { {\displaystyle 
  \tr \langle U(-\infty-i\beta,t) U(t,0) U(0,-\infty)
 \rangle }}
  \;,
 \label{kappa_def}
\end{equation}
where $\beta\equiv 1/T$; 
$gE_i \equiv i [D_0,D_i]$ is the color-electric field; 
$U(t_b,t_a)$ represents a temporal Wilson line from $t_a$ to $t_b$
at a fixed spatial location $\vec x=\vec{0}$; and the trace is over
the fundamental representation.  
The denominator removes any regularization issues
associated with the Wilson lines themselves.
The zero-frequency limit $\kappa(0)$ of this correlator
is the momentum diffusion coefficient of a heavy
quark~\cite{ct,eucl}, and the combination
$\eta_D = \kappa(0) / 2 M_\rmi{kin}T$ emerging from linear 
response relations, with $M_\rmi{kin}$ denoting  the so-called
kinetic mass of the heavy quark, determines the heavy quark 
thermalization rate~\cite{mt,eucl,bs}.  

Our goals and the organization of the paper are as follows.
In \se\ref{se:idea}, we describe the basic ideas behind classical 
lattice gauge theory as a tool for studying real-time quantities in QCD.
In \se\ref{se:observable}, we focus more precisely on the observable
in \eq\nr{kappa_def}, and use the weak-coupling regime to study how 
close the analogy between the two theories 
really is. The limitations of classical lattice 
gauge theory as a model for QCD are also illustrated, by studying the 
unphysical ``strong-coupling'' limit of the lattice-regulated theory.
In \se\ref{se:results}, we discuss the results we obtain at
intermediate couplings, where the weak-coupling expansion fails yet 
classical lattice gauge theory still captures the correct infrared
dynamics that causes the failure. Some discussion and conclusions
can be found in \se\ref{se:concl}, while two appendices contain
details related to the weak and strong-coupling regime on 
the classical lattice, respectively.

%
\section{Classical Lattice Theory:  Basic Idea}
\la{se:idea}

At a temperature far above the confinement scale, such 
that the effective gauge coupling $g$ is small,  QCD 
 (whether pure-glue or with dynamical quarks)
possesses
three different parametric length scales ($\equiv$ inverse momentum scales):
\bi
\item
the length scale $(\pi T)^{-1}$, where most of the energy resides;
\item
the ``color-electric'' length scale $(gT)^{-1}$, where plasma screening 
effects become important and perturbation theory needs to 
be resummed \cite{BraatenPisarski}; and
\item
the ``color-magnetic'' length scale $(g^2 T/\pi)^{-1}$, 
where interactions become genuinely non-perturbative. 
The longest spatial correlation lengths of gauge invariant 
operators are on this scale~\cite{linde}.
\ei
Different real-time correlation functions and physical properties of QCD are 
sensitive to different scales.  For instance, the Chern-Simons diffusion 
rate is sensitive dominantly to the momentum scale $g^2 T/\pi$; it depends 
on the other scales only in that they change the dynamics on this scale.  
Scattering, radiation and energy loss are sensitive 
mostly to the scale $gT$.  This scale therefore captures 
much of the physics of current interest in heavy ion collisions,
such as jet quenching and heavy quark thermalization.  Shear viscosity, 
on the other hand, is principally sensitive to the scale 
$\pi T$, since most of the energy and momentum reside there.

Since for the momentum scales $p\sim gT, g^2T/\pi$ the loop
 expansion parameter related to bosonic fields, 
$\epsilon \sim g^2\hbar/(e^{\beta \hbar  p} -1) $, 
can parametrically be replaced with its classical limit, $g^2T/p$,
it can be argued~\cite{oldclas} and shown 
formally~\cite{hbarsquared} that the physics at the scales $gT$,
$g^2 T/\pi$ is described by classical statistical field theory.
Quantum mechanics is only relevant at the scale $\pi T$, where its role is to
ensure that thermal excitations on short scales are suppressed.  The idea 
of the lattice analogue theory is to suppress thermal excitations on 
short scales instead by imposing a spatial lattice cutoff.  
The resulting theory is a classical field theory on all scales. At weak 
coupling it remains a three-scale theory, with:
\begin{itemize}
\item
the length scale $a$, where most of the energy resides;
\item
the color-electric length scale $(g^2 T/a)^{-1/2} \sim a \betaL^{1/2}$, 
where plasma screening effects become important and perturbation theory 
needs to be resummed \cite{bms,Arnold}; and
\item
the length scale $(g^2 T/\pi)^{-1} \sim a \betaL$, 
where interactions become non-perturbative.  
The longest spatial correlation lengths of gauge invariant
operators are on this scale.
\end{itemize}
Here we have introduced  the ``lattice coupling'', 
$
  \betaL \equiv {2\Nc}/{g^2 Ta} 
$,
which controls whether interactions are 
perturbative at the lattice spacing scale.

Physics at the scale $a$ is definitely different 
from physics of the quantum theory at the scale $(\pi T)^{-1}$.  
In particular, lattice discretization breaks
translational invariance so there is no conserved momentum.  This changes
hydrodynamic behavior in an essential way, so one should not try to study
shear viscosity with the classical theory.  
However, the more infrared scales, describing ``collective phenomena'', 
are only changed to the extent that the loop effects they feel from 
the hard momenta, 
associated with the expansion parameter $\epsilon \sim g^2 T/p \sim g^2 T a$,
differ from the corresponding 
quantum loop effects, with $\epsilon \sim g^2/\pi$.

For equal-time quantities,
these radiative effects turn out to be rotationally invariant and of exactly
the same form as in the continuum quantum theory.  They are simply a Debye
mass parameter for the $A_0$ field, of magnitude \cite{FKRS}
\begin{eqnarray}
\label{md_contin}
 m^2_{\rmi{D,cont}} & = & \frac{2 \Nc + \Nf}{6} g^2 T^2 \, , 
 \\
 m^2_{\rmi{D,latt}} & = & \frac{2\Nc \Sigma}{4\pi} \frac{g^2 T}{a} \,,
 \qquad
 \Sigma=3.175911536\ldots \,.
\label{md_latt}
\end{eqnarray}
Equating these gives a concrete way of relating the lattice spacing $a$ and
the temperature $T$, $a\simeq 3\Sigma\Nc/\pi T(2\Nc+\Nf)$.
  [Note that the $p\sim 1/a$ lattice modes are playing the role both
  of ultraviolet gluonic degrees of freedom (the $2\Nc$) as well 
  as of quark fields (the $\Nf$); 
  infrared quarks can be neglected because at
  low frequencies the Fermi-Dirac distribution function is much smaller
  than the Bose-Einstein one.]
At the dynamical level, however, 
radiative effects are no longer rotationally invariant~\cite{bms,Arnold}, 
which means that any
``matching'' between the lattice scale $a$ and the temperature $T$ is 
ambiguous, and only makes sense in order-of-magnitude.  
We return to this issue in more detail in the next section. 

Considering finally the color-magnetic scale, it remains 
the same, $g^2T/\pi$, in both theories. In other words, 
the precise form of the ultraviolet regulator is invisible
to physics at the largest distances. 
At the same time, the dynamics on the scale $g^2T/\pi$ 
is non-perturbative~\cite{linde}, but can relatively easily be 
simulated numerically through the classical description. 

Consider now the ``weak-coupling regime'' where
the three scales are widely separated, and an observable 
dominantly determined by the scale $gT$. Its exact value is given 
by the leading order result modified by relative corrections 
suppressed by $\epsilon \sim g^2 T/p$. While corrections from 
the hard scale $p\sim \pi T$ may remain controllably small down
to low temperatures (cf.,\ e.g., ref.~\cite{gE2}),
experience with 
many observables such as the plasmon frequency~\cite{plasmon}, 
the heavy quark diffusion coefficient~\cite{CaronMoore}, 
the light quark dispersion relation~\cite{fermion_mass},
the jet quenching parameter $\hat{q}$~\cite{Caron_qhat}, 
or the Debye screening length~\cite{mDebye}, 
has shown that  radiative
corrections from the scale $p\sim gT$ itself
%
%
can be very large (even if parametrically perturbative). 
The challenge would therefore be to sum the 
corrections from the scales $p\sim gT, g^2T/\pi$
to all orders. 

The key property of the classical lattice theory is that 
it is amenable to a numerical simulation, and therefore 
indeed allows for all-orders resummations of the type mentioned to be 
carried out in practice. Should the large corrections come from 
the scale $p\sim gT$ in the quantum theory, they are somewhat 
distorted in the classical lattice gauge theory, but the 
results are still representative of the qualitative behavior. 
%
%
The contact
to the quantum theory is only lost in the ``strong-coupling regime''
where $\betaL \lsim 1$; then all three scales are of the same order
and the physics differs essentially from the quantum theory.

%
\section{Classical Lattice Theory: Electric Field Correlator}
\la{se:observable}

In the previous section we argued on general grounds that 
classical lattice gauge theory and thermal QCD have qualitative and, 
in the very infrared, even quantitative similarities. We now 
want to demonstrate this explicitly for the case of
heavy quark thermalization. 

Let us start by recalling 
the reason for why the zero-frequency limit of \eq\nr{kappa_def} 
describes heavy quark momentum diffusion~\cite{ct} and consequently 
thermalization~\cite{mt,eucl,bs} 
(on the formal level, the correspondence
can be derived 
by making use of the heavy quark effective field theory~\cite{eucl}). 
In a classical framework,
it is quite easy to see why this is the case.  
Intuitively, the field $gE_i$ exerts a Lorentz force 
on a charge carrier, and $\kappa \equiv \kappa(0)$ is the total
correlation of that force with previous and future forces.  
The force changes the momentum of the heavy quarks. 
The classical lattice theory does not contain any heavy quarks,
but we can still evaluate the force--force correlation
function to see what momentum diffusion a  heavy quark
would feel. 

We discretize classical lattice gauge theory as in
ref.~\cite{Ambjornetal} and sample its thermal ensemble using the algorithm 
of ref.~\cite{mu_NCS} (we recommend ref.~\cite{mu_NCS}
for a more detailed description 
of the procedure). The classical simulation is generally carried out
in a gauge where the temporal links equal unity; thereby all the Wilson
lines disappear from \eq\nr{kappa_def}, and we only need to correlate
the electric fields. For a comparison we have also carried out some 
simulations with the ``improved'' lattice action of 
ref.~\cite{imp_action}, which provides a dispersion relation conforming 
more tightly to the continuum one, though it also differs significantly 
for $p\sim 1/a$. On the quantum theory side operator ordering plays a role;
in the following we assume symmetric ordering (for details, 
see ref.~\cite{eucl}), whereby the quantum correlator has the 
same symmetries as the classical one; this can be obtained by  
$\kappa(\omega) \rightarrow [{\kappa(\omega)+\kappa(-\omega)}]/{2}$ from
the case literally shown in \eq\nr{kappa_def}.

%
\begin{figure}[bt]
\centerline{%
   \epsfysize=7.0cm\epsfbox{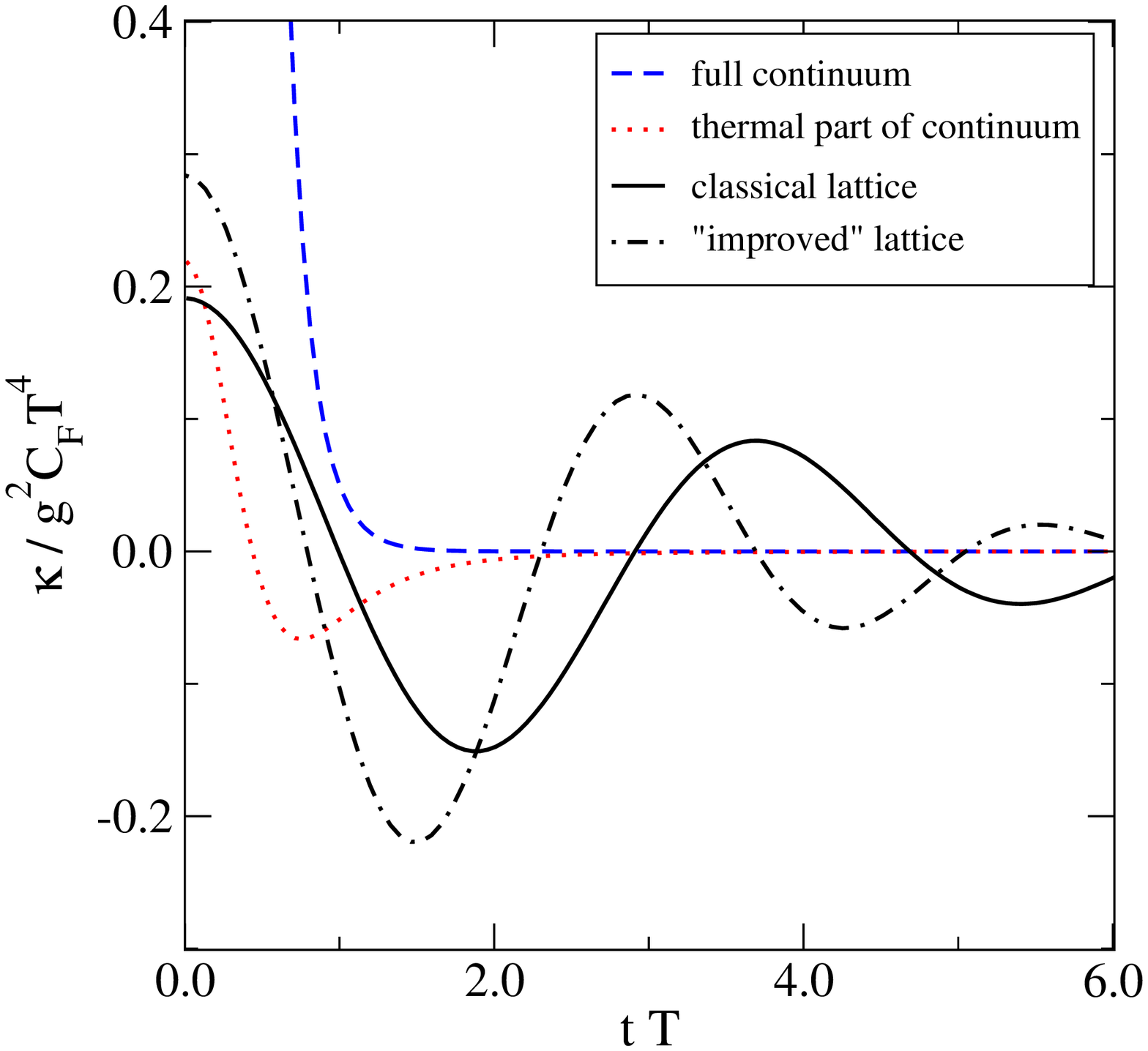}
~~~\epsfysize=7.0cm\epsfbox{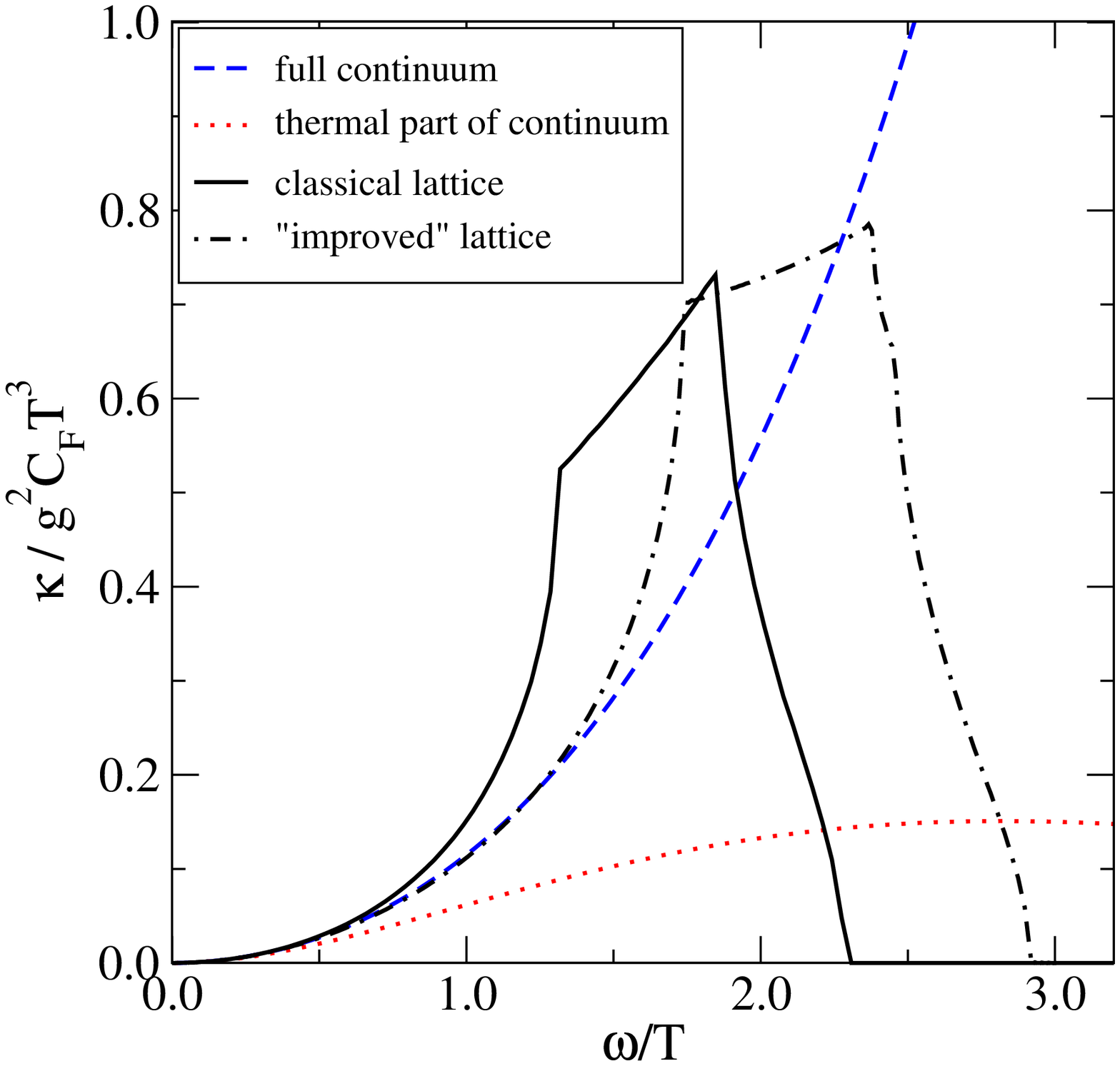}
}
\caption[Free correlation function]
{\label{fig_leadorder}\small
The correlation function $\kappa(t)$ (left) and its Fourier transform
$\kappa(\omega)$ (right) in the quantum continuum and classical lattice
theories, at leading order (free level). To relate the theories, we convert 
the lattice spacing $a$ to the inverse temperature $1/T$ by equating Debye
lengths for the pure-glue theory ($\Nf = 0$), whereby $a = 3 \Sigma/2\pi T$
(cf.\ discussion below \eq\nr{md_latt}).
}
\end{figure}
%

To start with we compare the theories at the free level. In the 
continuum theory, 
\ba
 \kappa_\rmi{cont}(\omega) & = &  [1 + 2 \nB{}(\omega)] 
 \frac{g^2 \Cf \omega^3}{6\pi}
 \;,\\ 
 \kappa_\rmi{cont}(t) & = &  
 g^2 \Cf  T^4 \pi^2 \biggl[ \frac{\cosh^2(\pi tT)}{\sinh^4(\pi tT)} 
 - \frac{1}{3\sinh^2(\pi tT)} \biggr]
 \;, 
\ea 
where $\nB{}$ is the Bose-Einstein distribution function. The vacuum 
behavior of $\kappa_\rmi{cont}(t)$ reads 
$g^2 \Cf  T^4 \pi^2 /(\pi tT)^4$.
In the classical lattice theory, on the other hand, 
\ba
 \kappa_\rmi{latt}(\omega) & = &
 \frac{2\pi g^2 \Cf T}{3}
 \int_{-\pi/a}^{\pi/a}
 \! \frac{{\rm d}^3\vec{p}}{(2\pi)^3}
 \Bigl[\delta(\tilde{p} - \omega) + \delta(\tilde{p} + \omega) \Bigr]
 \;,\la{k_l_w} \\ 
 \kappa_\rmi{latt}(t) & = &  
 \frac{2 g^2 \Cf T}{3}
 \int_{-\pi/a}^{\pi/a}
 \! \frac{{\rm d}^3\vec{p}}{(2\pi)^3}
 \cos(\tilde{p} t)
 \;, 
\ea
where $\tilde{p} \equiv \sqrt{\tilde{p}^2}$, 
$\tilde{p}^2 \equiv \sum_i \tilde{p}_i^2 \equiv 
\sum_i (\fr{2}{a} \sin\fr{ap_i}{2})^2$.
The results are plotted%
\footnote{%
    The ``thermal part''
    is the difference of the full and vacuum parts, 
    and it is this difference which is relevant
    for heavy quark thermalization.}
in \fig\ref{fig_leadorder}.  At first sight the
lattice correlator and the vacuum-subtracted thermal correlator
do {\em not} look alike:
the main difference is in the large-time behavior, where the
continuum correlator dies away but the lattice correlator
displays decaying oscillations. The difference is explained when we look at the
frequency-domain correlation functions.
Here we see that the lattice correlator has cusps while the continuum
correlator is smooth. The cusps are van Hove singularities which arise
because the lattice excitations follow 
a modified dispersion relation\footnote{%
  The dispersion relation for the ``improved'' action is more
  complicated, see \eqs(66,67) of ref.~\cite{imp_action} 
  [\eqs(63,64) in the journal version];
  the overall sign is wrong in the latter equation.
 },
\begin{equation}
 \label{eq:dispersion}
 \omega^2_{\rmi{latt}}(p) = 
 \tilde{p}^2 \,,
\end{equation}
which has vanishing slope at the corners of the Brillouin
zone, $\vec{p} = (n_1,n_2,n_3)\pi/a$, leading 
to cusps in the density of states at $\omega^2_\rmi{latt}=(4,8,12)/a^2$.
These van Hove singularities are well understood and have
little impact on the small-$\omega$ behavior which is actually of
interest.  In the small-frequency region $\omega \ll T$
the theories do agree completely.
In other words, the rather dramatic difference in the time-domain 
behavior shown
on the left in \fig\ref{fig_leadorder} arises because $\kappa(t)$ is
principally sensitive to the $T$ or $1/a$ scale, where the theories are
different; when we look at the frequency domain, the large frequency behaviors
are very different as expected, but the low frequency parts agree.
However, at the free level, the intercept $\kappa(0)$ is zero on both sides, 
so we need to consider interactions.

The leading non-zero value for $\kappa(0)$ turns out to involve
a logarithm of the scales $\pi T$ and $gT$. Considering the logarithm from the
IR side, its origin lies in the fact that the electric gauge field self-energy 
gets an imaginary part for $|\omega| < |\vec{p}| \sim gT$, corresponding to 
the phenomenon of Landau damping (it also gets a real part, corresponding
to Debye screening). The result for $\kappa$ is related to the cut 
of the electric field propagator, and can be written as
\ba
 \kappa_\rmi{cont} & \simeq &
 \frac{8\pi g^4 C_F \Nc}{3}
 \int_{p\ll  T} 
 \! \frac{{\rm d}^3 \vec{p}}{(2\pi)^3} \frac{{p}^2}{(p^2 + \mD^2)^2}
 \int \! \frac{{\rm d}^3 \vec{q}}{(2\pi)^3}
 \, \delta(({p-q})^2 - {q}^2) \, 
 q \, \nB{}(q) [1 + \nB{}(q)] 
 \nn & \simeq &
 \frac{g^2 C_F T \mD^2}{6\pi}  
 \biggl(\ln\frac{T}{\mD} + ... \biggr)
 \;,   \la{kappa_log_cont}
\ea
where $C_F \equiv (\Nc^2-1)/2\Nc$
is the Casimir of the heavy quark
representation.  Here we omitted for notational
simplicity quarks, and carried out the integral 
$
 2 g^2 \Nc \int_{\vec{q}} 
 \nB{}(q) [1 + \nB{}(q)] 
 = T \mD^2 
$.
The full computation and the result for the coefficient 
accompanying the logarithm can be found in ref.~\cite{mt}.

Consider then the classical lattice theory side.
Restricting again to the leading logarithmic order, the only
modifications needed are as follows: 
\bi
\item
 The statistical functions $\nB{}(q)$ and $1+\nB{}(q)$ are replaced by their
 classical limits $T/q$;
\item
 Dispersion relations and propagators use 
 $\tilde{p}_i$, $\tilde{q}_i$ in place of $p_i$, $q_i$.
\ei
Furthermore, 
at the order considered, the diagram is dominated by small exchange 
momentum ${p}$, and we can approximate the argument of the
$\delta$-function as 
$
 (\widetilde{p-q})^2 - \tilde q^2 =
 2 \tilde{\vec{p}}\cdot\ring{\vec{q}}
 + \rmO(p_i^2)
$, 
where $\ring{q}_i \equiv \fr{1}{a}\sin(aq_i)$.
Thereby the lattice version of \eq\nr{kappa_log_cont} becomes
\ba
 \kappa_\rmi{latt} & \simeq &
 \frac{4\pi g^4 T^2 C_F \Nc}{3}
 \int_{-\pi/a}^{\pi/a} 
 \! \frac{{\rm d}^3 \vec{p}}{(2\pi)^3} 
 \frac{\tilde{p}^2}{(\tilde{p}^2 + m^2_{\rmi{D,latt}})^2}
 \int_{-\pi/a}^{\pi/a} \! \frac{{\rm d}^3 \vec{q}}{(2\pi)^3}
 \frac{ \delta( \tilde{\vec{p}}\cdot\ring{\vec{q}} )}{\tilde{q}}  
 \nn & \simeq &
 \frac{g^4 T^2 C_F \Nc}{3\pi} 
 \int_{-\pi/a}^{\pi/a} \! \frac{{\rm d}^3 \vec{q}}{(2\pi)^3}
 \frac{1}{\sqrt{\tilde{q}^2\ring{q}^2}} 
 \times \biggl(\ln\frac{1}{a m_{\rmi{D,latt}}} + ... \biggr)
 \;.   \la{kappa_log_latt}
\ea
Here we made use of the fact that for $p\sim m_{\rmi{D,latt}} \ll 1/a$, 
$\vec{p}$ can be viewed as a continuum variable, so one can carry out
an angular integral to remove the $\delta$-function. 
We calculate the constant accompanying the logarithm, 
  denoted by $...$ in \eq\nr{kappa_log_latt}, 
in Appendix~A, finding it to be 1.8313(2).

In \eqs\nr{md_contin}, \nr{md_latt}, we saw the correspondence 
\be
 m^2_{\rmi{D,cont}} \leftrightarrow 2 g^2\Nc T \frac{\Sigma}{4\pi a}
 \;, \la{rel1} 
\ee
where the defining expression for $\Sigma$ reads
\begin{equation}
 \frac{\Sigma}{4\pi a} \equiv
 \int_{-\pi/a}^{\pi/a} \! \frac{{\rm d}^3 \vec{q}}{(2\pi)^3}
 \: \frac{1}{\tilde q^2} \,.
\end{equation}
Comparing the coefficients of the logarithms in 
\eqs\nr{kappa_log_cont}, \nr{kappa_log_latt}, on the other hand, 
suggests the correspondence 
\be
  m^2_{\rmi{D,cont}} \leftrightarrow 2 g^2\Nc T 
 \int_{-\pi/a}^{\pi/a} \! \frac{{\rm d}^3 \vec{q}}{(2\pi)^3}
 \frac{1}{\sqrt{\tilde{q}^2\ring{q}^2}} 
 \;. \la{rel2}
\ee 
The difference between \eqs\nr{rel1}, \nr{rel2} 
is a manifestation of the ambiguity in the matching of the continuum 
and lattice theories that was mentioned in \se\ref{se:idea}. More generally, 
noting that the dispersion relation in \Eq{eq:dispersion} 
gives a group velocity
\begin{equation}
|v_{\rmi{group}}(q)| \equiv \sqrt{ \frac{ \ring{q}^2}{\tilde{q}^2}}\,,
\end{equation}
and defining
\begin{equation}
 \frac{\Sigma_{v^n}}{4\pi a}
 \equiv 
 \int_{-\pi/a}^{\pi/a} \! \frac{{\rm d}^3 \vec{q}}{(2\pi)^3}
 \: \frac{|v_{\rmi{group}}(q)|^n}{\tilde q^2} 
 \;, 
 \label{eq:Sigmavi}
\end{equation}
we can write
\begin{equation}
 \kappa_\rmi{latt} \approx 
 \frac{g^2 \Cf T m_\rmi{D,latt}^2}{6\pi} \frac{\Sigmavi}{\Sigma} 
 \times \biggl(\ln\frac{1}{a m_{\rmi{D,latt}}} + 1.831 \biggr)
 \,. \la{kappa_latt}
\end{equation}
A similar correspondence  was 
found in ref.~\cite{Arnold} for a number of other quantities:  
scaling away $m_\rmi{D,latt}^2$, 
the Debye screening length involves
$\Sigma_{v^0} = \Sigma$; infrared magnetic damping involves
$\Sigma_{v^1}$; and the plasmon oscillation frequency involves
$\Sigma_{v^2}$.%
\footnote{%
    Each $v$ dependence arises from simple physics.  $\kappa_\rmi{latt}$ is
    the mean squared momentum a charged particle absorbs due to Coulomb
    interactions with passing excitations.  An excitation with velocity $v$
    has a flux factor suppressed by $v$ but it interacts for $1/v$ times as
    long, giving a force-squared enhanced by $v^{-2}$; hence
    $\kappa_\rmi{latt}$ involves $v^{-1}$.  Magnetic damping is similar but it
    involves magnetic forces, which are suppressed by $v$ relative to
    Coulombic forces.  Therefore the mean squared momentum exchange scales
    as $v^{0}$, leaving only the flux factor $v^1$.  Plasma oscillations
    involve the mean squared current generated by an oscillating
    electric field; the current is proportional to $v$ of the charges,
    so $\omega_\rmi{pl}^2 \sim v^2$.  Debye screening is a thermodynamic
    property so it shows no scaling with group velocity. }  
What we have shown is that Coulombic scattering
(electric damping)
involves $\Sigmavi$.  The numerical value of each $\Sigma_{v^n}$ is
given in Table \ref{table:sigma}.  The fact that they do not coincide means
that there is some ambiguity in how to relate
the lattice and continuum theories.
Nevertheless, 
the general structures of the answers 
in \eqs\nr{kappa_log_cont}, \nr{kappa_log_latt}, 
including the existence of logarithms, 
are the same, whereby we can conclude that the dynamics
of the two theories indeed bear a strong {\em qualitative} 
resemblance to each other. 

%
\begin{table}
\centerline{\begin{tabular}{|c|l|l|} \hline
 & ``standard''\cite{Ambjornetal} & ``improved'' \cite{imp_action} \\ \hline
$\Sigma_{v^2}$ & 1.6222746498   & 1.78576519  \\
$\Sigma_{v^1}$ & 2.1498783949   & 2.13792379  \\
$\Sigma_{v^0}$ & 3.1759115356   & 2.783189232 \\
$\Sigmavi$     & 5.5079614967   & 4.1679252   \\ \hline
\end{tabular}}
\caption{\small 
  Values of $\Sigma_{v^n}$ for the standard and ``improved'' lattice
  actions.  The spread of values is a measure of how different the structure
  of the lattice Hard Thermal Loops is from the continuum ones.
\label{table:sigma}}
\end{table}
%

The discussion so far has assumed that we are in the weak-coupling
regime, i.e.\ that $\betaL$ is large. If $\betaL$ decreases so much that 
all three momentum scales are of the same order of magnitude, then 
the behavior of $\kappa_\rmi{latt}$ changes significantly. Scaling 
$\kappa_\rmi{latt}$ dimensionless by multiplying with $a^3$, and 
making use of the definition of $\beta_L$, the weak-coupling 
behavior in \eq\nr{kappa_latt} corresponds to 
\be
 a^3 \kappa_\rmi{latt} \; \stackrel{\betaLl \gg 1}{\sim} \; 
 \frac{1}{\betaL^2} \ln \betaL
 \;, 
\ee
while at small $\betaL$ we find, through the arguments in 
Appendix~B, the behavior 
\be
  a^3 \kappa_\rmi{latt} \; \stackrel{\betaLl \ll 1}{\sim} \; 
  \frac{1}{\betaL^{5/2}}
 \;. \la{strong_coupl}
\ee
The physics behind this functional form is very specific to 
the nature of the lattice variables, however, so we do not 
expect any analogy with the continuum quantum theory in the latter 
regime, and refrain from a further discussion here.

%
\section{Numerical Results}
\la{se:results}

In the previous section we have verified that, 
at weak coupling, the electric field correlator
in classical lattice gauge theory behaves quite 
similarly to physical QCD: 
if we fix the lattice spacing $a$ by equating the Debye
screening lengths, the leading-logarithmic $\kappa$ 
of the classical lattice theory is larger
than that in the quantum continuum theory by a factor
$\Sigmavi/\Sigma_{v^0} \sim 5/3$. We now proceed to 
larger values of the coupling (smaller $\betaL$)
with the help of numerical simulations.  
At relatively small coupling (large $\betaL$), 
we can check how fast the weak-coupling regime is approached. 
At stronger coupling (intermediate $\betaL\sim 1$), 
we can find out whether the leading-order weak-coupling
result is reasonable even in order of magnitude, and 
whether it  underestimates or overestimates the actual behavior.  
This then gives us some guidance for what to expect in QCD.

%
\subsection{Intercept at $\omega\to 0$}

\begin{figure}[t]


\centerline{%
   \epsfysize=7.0cm\epsfbox{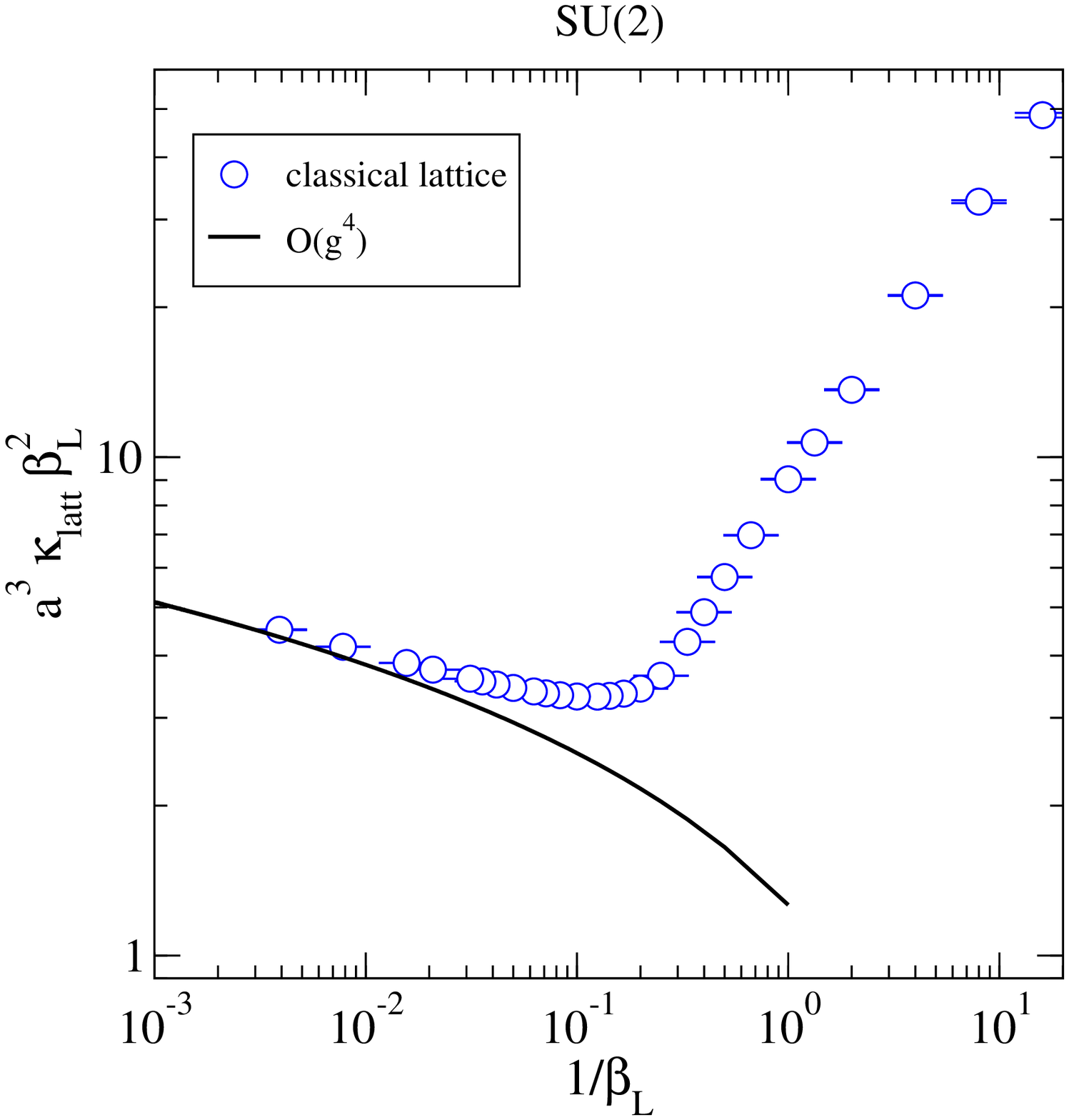}%
~~~\epsfysize=7.0cm\epsfbox{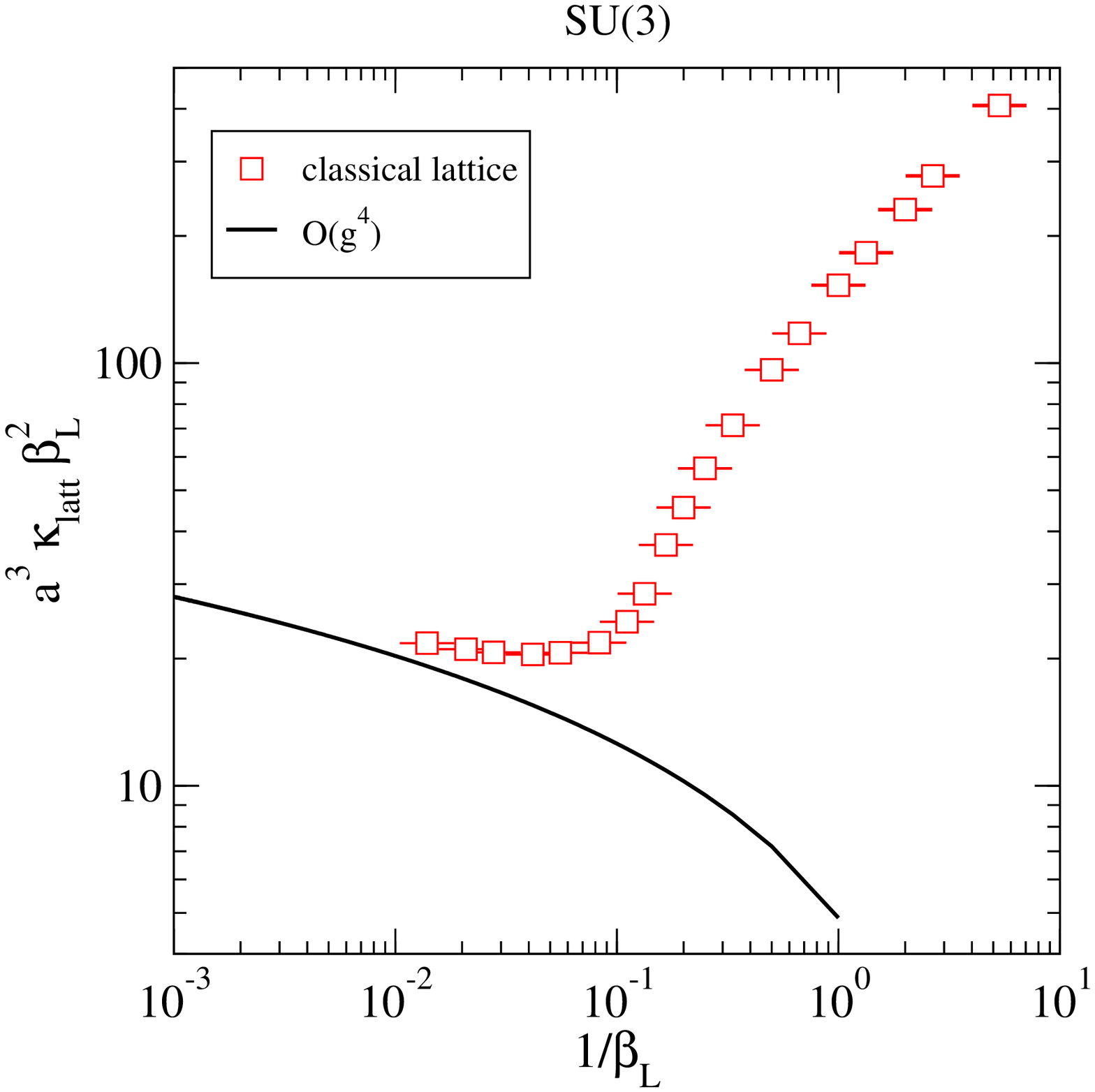}%
}

\caption[a]{\small 
  Numerical results for the intercept $\kappa_\rmi{latt}$ (open symbols), 
  compared with the weak-coupling prediction from \eq\nr{kappa_latt}
  (line). The left plot is for SU(2), the right one for SU(3). 
  Note that $1/\beta_L = g^2 T a/2\Nc$ scales like $\alpha_s$, assuming
  the matching $a \sim 1/T$ (cf.\ discussion after \eq\nr{md_latt}).
 }
\la{fig:intercept}
\end{figure}

Our numerical results for $\kappa_\rmi{latt}$, compared with the 
leading-order weak-coupling result, are shown in \fig\ref{fig:intercept}, 
both for SU(2) [included because a large $\betaL$-range  could
be scanned with a modest numerical effort] and for SU(3).\footnote{%
    In the numerical implementation the theory is discretized in time as
    well as space, but with a much finer spacing, and our numerical results
    for $\kappa_\rmi{latt}$ represent the limit of zero temporal spacing.
    We have also checked that our results contain no significant finite
    volume or non-zero $\omega$ artifacts. 
} 
We note, first of all, that at large $\betaL$, the results
approach the analytic ones of \eq\nr{kappa_latt}. However, 
as soon as $\betaL\lsim 100$, the non-perturbative results  
deviate from the leading-order ones. The non-perturbative results 
are always {\em larger} than the perturbative estimate. 
For $\betaL = 1 ... 10$, a crossover takes place\footnote{%
 We have checked that there 
 is no actual phase transition in the thermodynamics 
 of the system. 
 }
from one type of behavior to another. At $\betaL \ll 1$,
the results approach the behavior of \eq\nr{strong_coupl}.
(We have not worked out the numerical prefactor for \eq\nr{strong_coupl}, 
and hence do not show the corresponding curves 
in \fig\ref{fig:intercept}.) 

In order to make quantitative use of the numerical results, 
it is convenient to change the units of both axes. Recalling 
the definition of $m^2_{\rmi{D,latt}}$ from \eq\nr{md_latt}, 
we choose the variable 
$g^2\Nc T/m_\rmi{D,latt} = 2 \Nc (\pi/\Sigma\betaL)^{1/2}$
as the $x$-coordinate; this quantity is the ratio of the $g^2T$ to $gT$
scales and is therefore the expansion parameter for perturbation theory at the
scale $gT$.  We also divide $a^3\kappa_\rmi{latt}$
by the coefficient of the leading logarithm, 
$a^3 g^2 \Cf T m_\rmi{D,latt}^2/6\pi = 
 \Cf \Nc^3 \Sigma/ 3 \pi^2 \betaL^2$.
In these units, the lattice results have a direct counterpart
in the continuum theory. The weak-coupling regime
is plotted in the new units in \fig\ref{fig:series}. 

In the continuum theory, 
corrections of $\rmO(g^5)$ to $\kappa_\rmi{cont}$
have recently been determined~\cite{CaronMoore}, and it is now interesting
to compare the results.  According to ref.~\cite{CaronMoore}, 
\begin{equation}
\kappa_\rmi{cont} = \frac{g^2 \Cf T m^2_\rmi{D,cont}}{6\pi} \left(
  \ln \frac{T}{m_\rmi{D,cont} } 
 + C_\rmi{cont} + D_\rmi{cont} \frac{\Nc g^2 T}{m_\rmi{D,cont} }
  + \ldots
  \right) \;, 
\end{equation}
with $D_\rmi{cont} = 0.7767$.  
The physics giving rise to $D_\rmi{cont}$ involves only the
length scale $gT$ and should be reproduced on the lattice; however,
it depends on the structure of the Hard Thermal Loops in an essential way, so
the lattice value could differ by 
up to ${\cal O}(50\%)$ as discussed in the previous section. 
Still, this motivates a fit of the lattice
data to the form
\begin{equation}
 \kappa_\rmi{latt} = \frac{g^2 \Cf T m^2_\rmi{D,latt}}{6\pi}
 \left( \frac{\Sigmavi}{\Sigma} \ln \frac{1}{a m_\rmi{D,latt}} + C_\rmi{latt} +
 D_\rmi{latt} \frac{\Nc g^2 T}{m_\rmi{D,latt}} + \ldots \right) \,,
\label{kappa_series}
\end{equation}
where $C_\rmi{latt} = 1.831 \times 
\Sigma_{v^{-1}}/\Sigma = 3.176$.
%
%
The coefficient $D_\rmi{latt}$
can confirm the sign and approximate magnitude of 
$D_\rmi{cont}$.  It can also tell us about the next terms in the expansion, in
particular whether the $\rmO(g^5)$ 
calculation is an underestimate or an overestimate
of the real $\kappa$.  Note, however, that
the next term, of $\rmO(g^6)$,
would receive contributions not only from the scale $gT$ 
but also from the scale $\pi T$, 
so the relation between the lattice and continuum
theories becomes less precise at this order.

\begin{figure}[t]


\centerline{%
   \epsfysize=8.0cm\epsfbox{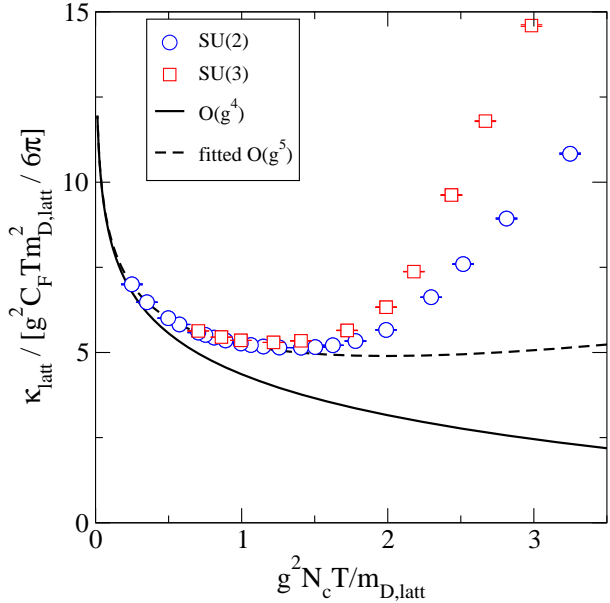}
}

\caption[NLO fit]{\small 
 $\kappa_\rmi{latt}$, normalized to the leading-order perturbative 
 behavior, expressed as a function of the expansion parameter  
 $g^2 \Nc T/m_\rmi{D,latt}$
 related to corrections from the scale $gT$.  
 At weak coupling, the SU(2) and SU(3) results agree and are
 well fit by the $\rmO(g^5)$ perturbative behavior.  At stronger coupling, 
 $\kappa_\rmi{latt}$ rises above the perturbative fit, 
 by a group dependent amount. Compared with \fig\ref{fig:intercept}, 
 the horizontal axis is restricted to $1/\betaL \le 0.77$ for SU(2), 
 $1/\betaL \le 0.34$ for SU(3).
 }
\la{fig:series}
\end{figure}

Fig.~\ref{fig:series} 
shows the (1-parameter) fit to the lattice data 
according to \eq\nr{kappa_series}. The fit is very good out
to $g^2\Nc T/m_\rmi{D,latt} \sim 1.5$. We extract the value
$D_{\rmi{latt}} = 0.87(4)$, in surprisingly good agreement with
$D_\rmi{cont}  = 0.7767$; and we see that the same coefficient $D_\rmi{latt}$ 
fits the SU(2) and SU(3) data, just as the continuum computation predicts.  
However, at larger couplings $\kappa_\rmi{latt}$ rises {\em above}
the fitted behavior, particularly for the group SU(3).

The above results no doubt depend on the details of the numerical
implementation of the lattice theory and of the electric field operator.
As a check on the robustness of our results, we re-compute them using
the ``improved'' lattice action of ref.~\cite{imp_action}.  This
action is tree-level improved so that the IR behavior naively
coincides more tightly with the continuum, as shown for instance by the
better IR behavior of the free-theory correlator $\kappa(\omega)$ in
\fig\ref{fig_leadorder}.  However the UV behavior still has (different)
anisotropic non-ultrarelativistic dispersion, so the Hard Thermal Loop
effects are not those of the continuum (though they are somewhat
closer, as reflected by the slightly narrower spread of the $\Sigma_{v^{n}}$
values in Table \ref{table:sigma}).  Therefore it is
better to think of this implementation as ``different'' rather than
truly ``improved.''

\begin{figure}[t]


\centerline{%
   \epsfysize=7.0cm\epsfbox{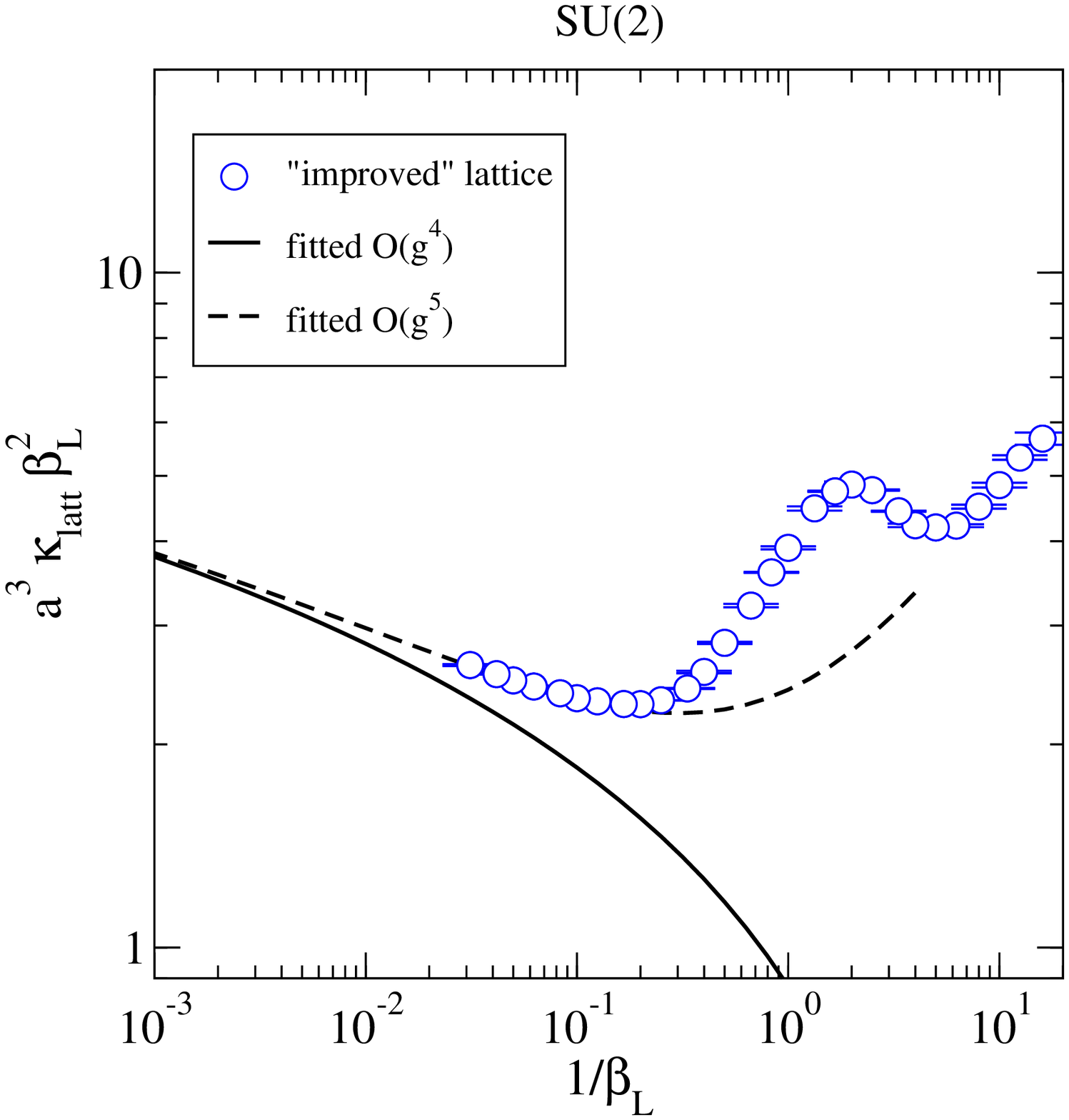}    
~~~\epsfysize=7.0cm\epsfbox{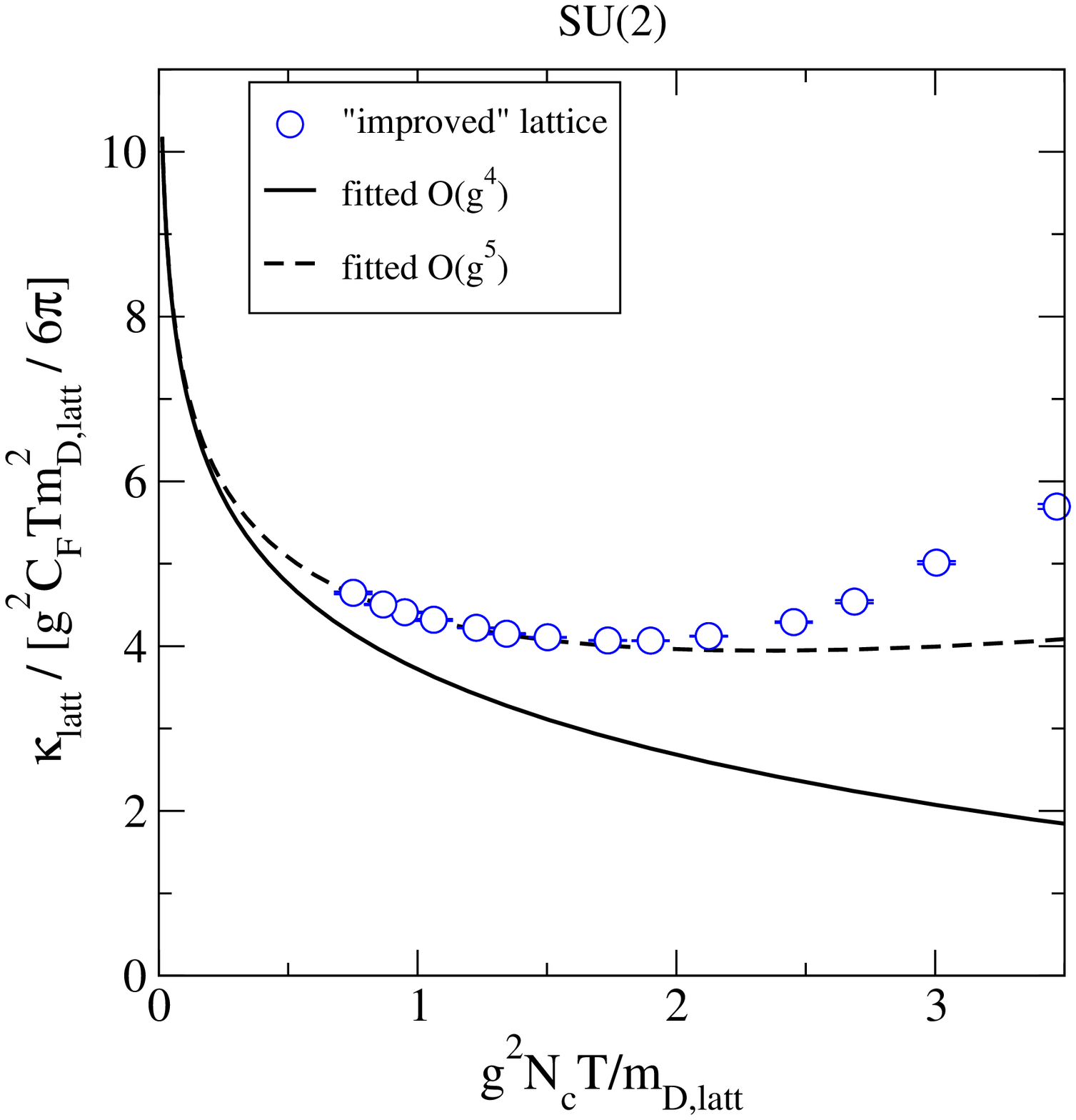}    
}

\caption[improved theory]{\small 
  $\kappa_\rmi{latt}$ using the ``improved'' lattice action.  
  Left: the overall behavior in lattice units. 
  Right: a magnification of the weak-coupling regime, normalized
  to the leading-order perturbative behavior.   The weak-coupling
  behavior is in good qualitative accord with the standard action, 
  but the strong coupling behavior is qualitatively different
  (cf.\ \figs\ref{fig:intercept}, \ref{fig:series}).
 }
\la{fig_imp}
\end{figure}

\fig\ref{fig_imp} shows $\kappa(0)$ as a function of the lattice
coupling for the ``different''/``improved'' 
lattice action for the group SU(2).  While
the lattice constants $C_\rmi{latt}=2.5$ and $D_\rmi{latt}=0.64$
(this time both are fitted) differ from 
the ``standard'' action values, the qualitative message is
the same; at weak coupling the behavior appears to be well described by
next-to-leading order perturbation theory, but at stronger coupling
perturbation theory is an underestimate.  But while for $\betaL\gsim 1$ the
two implementations give similar qualitative results, at extremely strong
coupling the behaviors are not even qualitatively the same.  This
reinforces our belief that the $\beta_\rmi{L} \lsim 1$ behavior is a lattice
artifact with no bearing on QCD.

We finally attempt a rough order-of-magnitude estimate for 
which $\betaL$-range corresponds to the situation met in heavy 
ion collision experiments. Combining the matching from below 
\eq\nr{md_latt} with the definition of $\betaL$, we get 
\be 
 \betaL \sim \frac{2\Nc + \Nf}{6 \Sigma \alpha_s}
 \;. 
\ee 
The relevant value of $\alpha_s$ can in the present context
probably best be approximated by taking it from the dimensionally reduced
effective theory~\cite{dr}, to which the classical lattice gauge
theory reduces in the case of equal-time observables. In this limit
the coupling has been 
computed up to 2-loop level~\cite{gE2}; for $\Nf=3$ the values
are $g^2 \sim 3 ... 2$ for $T/\Lambdamsbar \sim 1 ... 4$,  
corresponding to $\alpha_s = 0.24 ... 0.16$, and 
subsequently $\betaL = 2 ... 3$. This corresponds to 
$g^2\Nc T/m_\rmi{D,latt} = 4...3$.

Remarkably, in the range $\betaL = 2 ... 3$, the numerical SU(3)
values in \fig\ref{fig:intercept} exceed the weak-coupling result
by as much as an order of magnitude! Though these values of $\betaL$
are so small that the matching cannot be trusted on any kind of 
quantitative level, such a huge effect is still encouraging 
both from the experimental point of view~\cite{exp}, where 
the apparently very rapid thermalization of heavy quarks remains
a mystery, as well as from the point of view of following the 
suggestion of ref.~\cite{eucl} in order to measure $\kappa_\rmi{cont}$
with Euclidean lattice Monte Carlo methods. Indeed, there
may well be an exciting qualitative discovery to be made 
on the lattice. 

%
\subsection{General shape of the spectral function}

On the point of lattice Monte Carlo simulations, ref.~\cite{eucl}
argued that the Euclidean analogue of \eq\nr{kappa_def} leads to 
a correlator, denoted by $G_E(\tau)$, which has a non-trivial 
continuum limit and can be related to the intercept $\kappa_\rmi{cont}$
through standard relations.  Specifically, the task would be to 
invert the relation 
\be
 G_E(\tau) = 
 \int_0^\infty
 \frac{{\rm d}\omega}{\pi} \kappa_\rmi{cont}(\omega)
 \frac{\cosh \left(\frac{\beta}{2} - \tau\right)\omega}
 {\cosh\frac{\beta \omega}{2}} 
 \;. \la{int_rel} 
\ee
It is a problem, though, that strictly speaking the relation in  
\eq\nr{int_rel} is not invertible without further input. In practice, 
this means that a certain {\it Ansatz} (sometimes called a prior) is needed, 
which is then refined through the numerical data. For this reason, 
significant efforts have been devoted to analytic computations
of spectral functions in the presence of a spatial lattice, in the 
limit of a high temperature, for cases such as the 2-point correlator 
of the vector current of heavy quarks~\cite{free_spectral}.

\begin{figure}[t]


\centerline{%
   \epsfysize=7.0cm\epsfbox{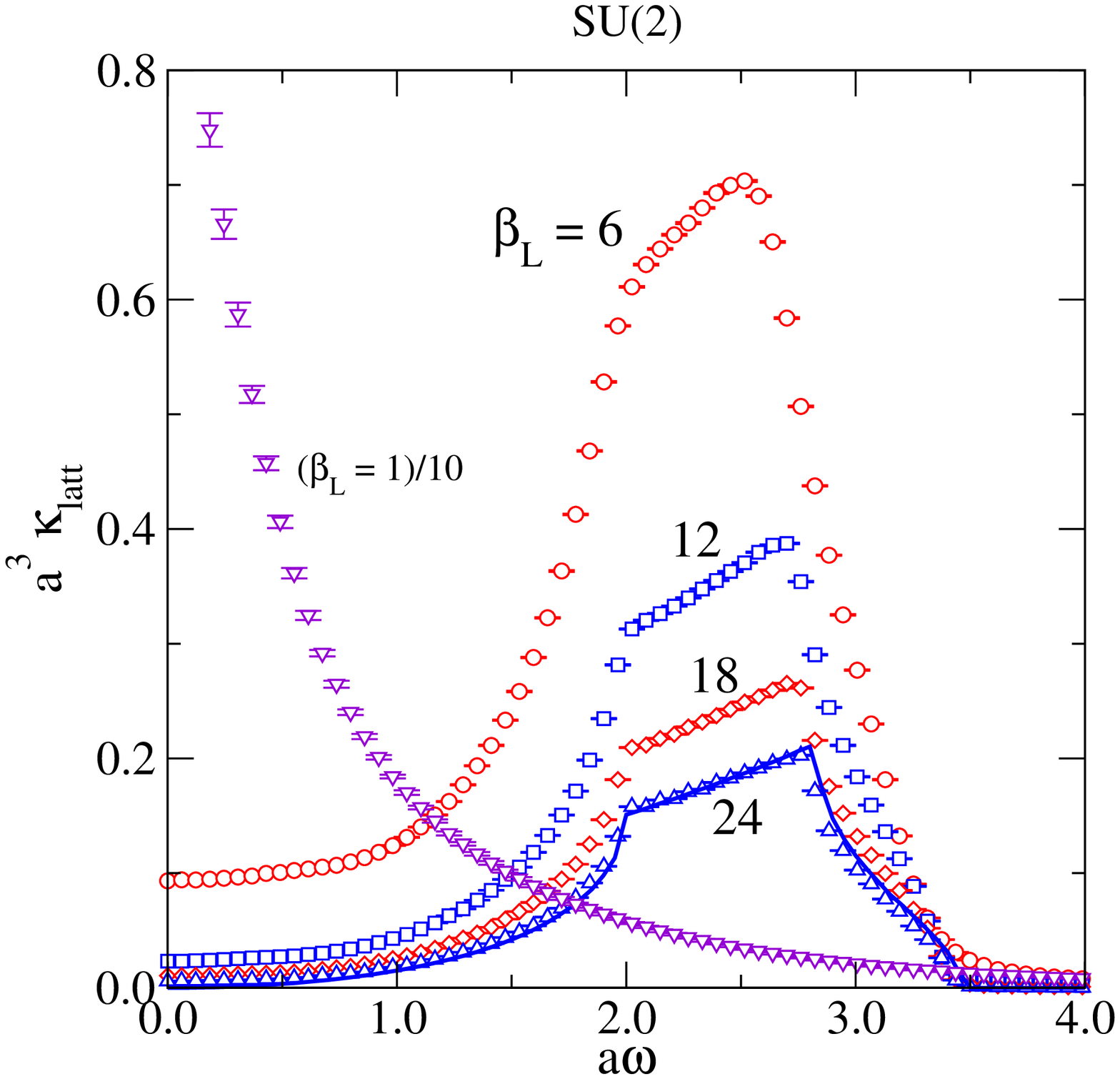}%
 ~~~\epsfysize=7.0cm\epsfbox{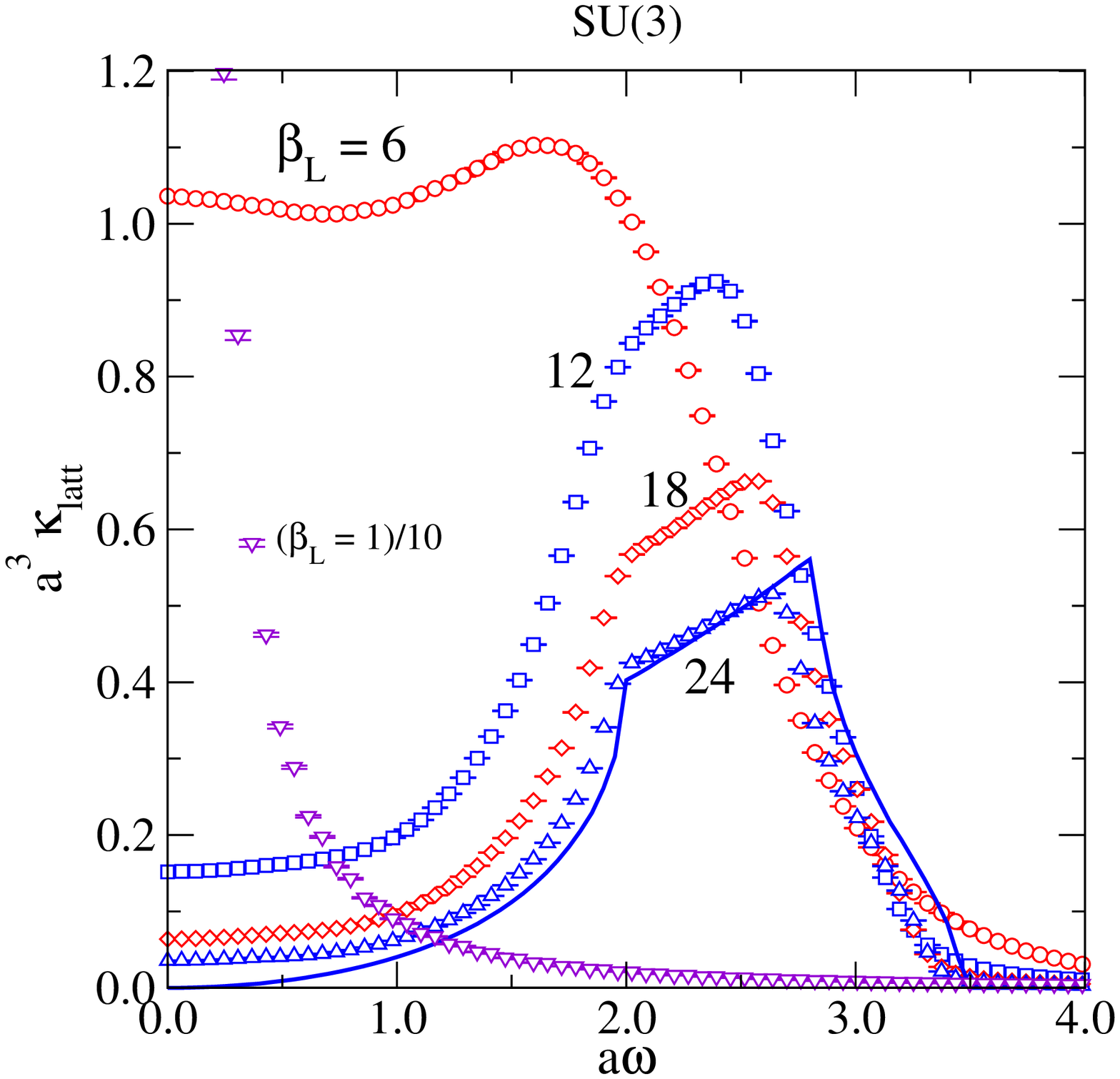}%
}

\caption[a]{\small 
  Numerical results for the function $a^3 \kappa_\rmi{latt}$ 
  (open symbols), 
  compared with the weak-coupling prediction from  \eq\nr{k_l_w} 
  for $\betaL=24$ (line). 
  The left plot is for SU(2), the right one for SU(3). For 
  the (unphysical) case $\betaL=1$, included as a reference 
  for the discussion in Appendix~B, we have divided the 
  central values (but not the error bars) by a factor 10.
 }
\la{fig:a3kappaw}
\end{figure}

We can now use our data, both perturbative as well as non-perturbative, 
to obtain an {\it Ansatz} for the spectral function 
$\kappa_\rmi{cont}(\omega)$. 
In \fig\ref{fig:a3kappaw}, results are shown for the function 
$a^3 \kappa_\rmi{latt}(\omega)$ at various $\betaL$, together with a comparison
with the free theory result. Noting that on the 4-dimensional lattice, 
$\beta = N_\tau a_\tau$, where $N_\tau$, $a_\tau$ are the number of 
lattice points and the lattice spacing in the time direction, respectively, 
  and naively enforcing the replacement of 
  the classical limit of the Bose-Einstein distribution
  function, $T/\omega$, by the corresponding quantum mechanical expression, 
  $1/2 + \nB{}(\omega)$, 
we can expect $\kappa_\rmi{cont}(\omega)$ to behave as 
\be
 {\kappa_\rmi{cont}(\omega)}
 \simeq
 \frac{\omega  a_\tau N_\tau}{2}
 \coth\biggl( \frac{\omega a_\tau N_\tau}{2} \biggr)
 \kappa_\rmi{latt}(\omega)
 \;. 
\ee
In particular, for $\omega\ll T$, $\kappa_\rmi{cont}(\omega)$ should be 
completely flat just like $\kappa_\rmi{latt}(\omega)$; 
moreover, in general, $\kappa_\rmi{cont}(\omega)$
should show no peaks other than 
at $\omega \sim (1.5 - 3.0)/a$, where $a$ is the spatial lattice spacing. 
We consider these qualitative features to be relatively 
robust, and they can in any case serve as crosschecks on 
particular practical inversions of \eq\nr{int_rel}.

Finally, we remark that the 
corresponding spectral functions computed for ${\mathcal N} = 4$
Super-Yang-Mills theory at infinite 't Hooft coupling in continuum 
show an analogous smooth behavior at small frequencies, taken over by 
ultraviolet physics at $\omega\sim T$~\cite{ct,ssg}.

%
\section{Summary and Outlook}
\la{se:concl}

The purpose of this paper has been to make use of classical 
lattice gauge theory, in order to gain insights on the dynamics
of QCD in the temperature range accessible to current and near-future
heavy ion collision experiments. We have stressed, in particular, 
that classical lattice gauge theory is a multiscale system 
just like QCD; unlike QCD, however, it easily lends itself to 
non-perturbative simulations of real-time observables, 
in both the weak-coupling and strong-coupling regimes. 
Thereby a semi-analytic understanding can be obtained of many
interesting observables, without changing the number of color degrees 
of freedom or introducing unphysical infrared fields. 

More specifically, we have elaborated on the heavy quark momentum 
diffusion coefficient, denoted by $\kappa$, 
which determines the heavy quark thermalization
rate through linear response relations. This quantity belongs to the 
general class of observables which are ``dominantly'' influenced by 
momenta around the Debye scale, $p\sim gT$. We have shown explicitly
through a weak-coupling analysis that while the physics of the 
classical lattice gauge theory differs from that in QCD on  
the quantitative level, 
by effects of up to 50\%, the qualitative features of the dynamics
do remain intact. 

Proceeding from weak coupling towards intermediate coupling, 
we have furthermore shown that the leading-order weak-coupling
expression, and 
even the larger next-to-leading order expression,
{\em underestimate} the non-perturbative result. Given 
the close analogy with QCD, the same statement should be true
on that side. This seems to give realistic hopes
that a future quantitative
determination of $\kappa$ through 4-dimensional 
lattice Monte Carlo simulations will reveal a large 
thermalization rate, which might help to explain the surprisingly
rapid thermalization that has been 
observed at RHIC experiments~\cite{exp}.

Finally, with regard to Monte Carlo simulations, we have explored
the general structure of the spectral function corresponding to the 
Euclidean electric field correlator that can be used for determining 
the thermalization rate~\cite{eucl}. We find that apart from 
a single peak at the scale of the spatial lattice spacing, 
the spectral function has little
structure, both at weak and at intermediate coupling. This should
be an encouraging message with respect to the analytic 
continuation needed in the analysis of the Monte Carlo simulations. 

%
\section*{Acknowledgments}

ML and GDM thank Simon Caron-Huot for valuable discussions. 
The work of GDM was supported in part by the Natural Sciences
and Engineering Research Council of Canada, and by the 
Alexander von Humboldt Foundation through an F.\ W.\ Bessel award.

\newpage


\appendix
\renewcommand{\thesection}{Appendix~\Alph{section}}
\renewcommand{\thesubsection}{\Alph{section}.\arabic{subsection}}
\renewcommand{\theequation}{\Alph{section}.\arabic{equation}}

%
\section{Weak-coupling regime in classical lattice gauge theory}
\label{app_A}

%
\begin{figure}[t]
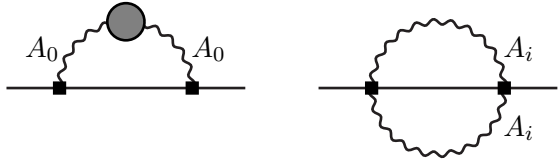


\begin{eqnarray*}
 \piccc{\Lsc(0,15)(90,15) %
 \PhotonArc(45,15)(25,0,180){1}{12}
 \GBoxc(20,15)(4,4){0}
 \GBoxc(70,15)(4,4){0}
 \GCirc(45,40){7}{0.5}
 \Text(20,30)[r]{$A_0$}
 \Text(70,30)[l]{$A_0$}
 } 
   \qquad 
 \piccc{\Lsc(0,15)(90,15) %
 \PhotonArc(45,15)(25,0,360){1}{24}
 \GBoxc(20,15)(4,4){0}
 \GBoxc(70,15)(4,4){0}
 \Text(70,30)[l]{$A_i$}
 \Text(70,0)[l]{$A_i$}
 }
\end{eqnarray*}

\vspace*{0.5cm}

\caption[Diagram for leading-order scattering]
{\label{diagram1}\small
The diagrams contributing to $\kappa_\rmi{latt}$ in Coulomb
gauge. The straight lines represent temporal Wilson lines, the closed
squares electric fields, and the grey bubble the 
gauge field self-energy.  
}
\end{figure}
%

We give in this appendix some details concerning the computation
of the constant accompanying the logarithm in 
the leading-order weak-coupling result, \eq\nr{kappa_log_latt}
(or \eq\nr{kappa_latt}). 

In the ``standard'' implementation~\cite{Ambjornetal} of 
classical lattice gauge theory, where time is continuous
and Minkowskian, the electric field strength has the form 
\be
 a E_i(x) = -\frac{i}{g} [\partial_t U_i(x)] U^\dagger_i(x) 
 + A_0(x) - U_i(x) A_0(x + a \hat i) U^\dagger_i(x) 
 \;. \la{Ei_def}
\ee
Here $U_i$ are the spatial link matrices. For a perturbative 
computation we write 
$
 U_i = \exp(i a g A^b_i T^b)
$, 
where $T^b$ are Hermitean and assumed normalized as 
$
 \tr[T^b T^c] = \delta^{bc}/2
$.
As usual~\cite{Rothe}, Fourier representations of the spatial 
variables are most conveniently chosen as 
$
 A^b_i(x) = \int_K  A^b_i(K) e^{i K\cdot(x + a \hat i / 2)}
$.
Furthermore a gauge needs to be fixed; like in the continuum 
computation~\cite{mt}, it is convenient to choose a Coulomb gauge 
so the propagator splits into a transverse spatial part and an $A_0$ field
propagator which has no on-shell spectral weight.
In this gauge the result emerges from
two graphs, depicted in \fig\ref{diagram1}: 
the $A_0$ self-energy diagram, mediated by the vertex
\be
 A^a_0(P)\, A^b_i(Q) \, A^c_j(R) \, \delta(P+Q+R)
 \frac{ig}{2} f^{abc} \, \delta_{ij} \cos({a p_i/2}) (R_0 - Q_0)
 \;,
\ee
plus from 
one additional graph, namely the bubble diagram sourced by the 
second term in the expansion of the first term of \eq\nr{Ei_def}, 
\be
 a E_i(x) = ... 
 + \fr12 a^2 g T^b f^{bcd}[\partial_t A^c_i(x)] A^d_i(x) +  ... \;. 
\ee 
The final result can be written as 
\ba
 \kappa_\rmi{latt} & = & 
 \frac{4\pi g^4 T^2 C_F C_A}{3}
 \int_{-\pi/a}^{\pi/a} \! 
 \frac{{\rm d}^3 \vec{p}}{(2\pi)^3} 
 \frac{\tilde{p}^2}{(\tilde{p}^2 + \mD^2)^2}
 \int_{-\pi/a}^{\pi/a} \! 
 \frac{{\rm d}^3 \vec{q}}{(2\pi)^3} 
 \frac{\delta((\widetilde{p-q})^2 - \tilde q^2)}{\tilde q}
 \nn & & \; \times
 \biggl\{ 
   2 - \frac{\tilde p^2}{\tilde q^2} 
 + \frac{\tilde p^4}{4\tilde q^4} + 
 \frac{a^2}{4}
 \sum_{i=1}^3
 \biggl[ 
  \frac{\tilde p_i^2 \tilde q_i^2 + 
  \tilde p_i^2 (\widetilde{p_i-q_i})^2}{\tilde q^2}
  + \frac{\tilde p^2 \tilde q_i^2 (\widetilde{p_i-q_i})^2 }{\tilde q^4}
 \biggr]
 \biggr\}
 \;, \la{kappa_full}
\ea
where the limit $\mD \ll 1/a$ is assumed. 

For the numerical evaluation of \eq\nr{kappa_full}, one can for instance
integrate explicitly over one of the momentum components, to remove the 
$\delta$-function, and carry out the remaining five-dimensional integral
numerically, simplifying the range by making use of various symmetries.  
More refined strategies are certainly possible but not necessary if 
only a few digits are needed.

%
\section{Strong-coupling regime in classical lattice gauge theory}
\label{app_B}

Figure \ref{fig:intercept} shows that, 
in the limit $\betaL \rightarrow 0$, the electric
field autocorrelator diverges as $\betaL^{-5/2}$, 
while \fig\ref{fig:a3kappaw} shows
that the frequency spectrum for $\kappa_\rmi{latt}(\omega)$ 
becomes tightly peaked at
small frequencies.  What is going on in this regime, and could it have
anything to do with QCD?  Here we show that 
the answer to the latter question
is almost certainly negative.

To do so we need to discuss a few features of the numerical 
simulation, which we have otherwise left to the references.  
Fixing to the temporal gauge, 
which is convenient because the temporal Wilson lines in the
definition, \Eq{kappa_def}, are identity operators,
the continuum Yang-Mills theory is described by gauge 
fields $A_i(\vec{x},t)$ and their canonical
momenta, the electric fields $E_i(\vec{x},t)$.  
On the lattice, the degrees of freedom are the dimensionless electric fields
${\cal E}_i = a^2 g E_i + \rmO(a^3)$ and the gauge links, 
$U_i(x) = \exp (iag A_i(x))$. We can write 
${\cal E}_i = {\cal E}_i^b T^b$, where   
$T^b$ are Hermitean generators of the SU($\Nc$) algebra, 
normalized as 
$
 \tr[T^b T^c] = \delta^{bc}/2
$.

The lattice simulation proceeds by sampling initial configurations
with a classical Hamiltonian, and then evolving the configurations 
in real time through classical equations of motion. 
The time evolution of the link matrices is
\begin{equation}
 a\, \partial_t U_i({x})  =  i \, 
 {\cal E}_i({x}) U_i({x}) 
 \;, \la{dU} 
\end{equation}
while for the electric fields it is 
\begin{eqnarray}
 a\, \partial_t {\cal E}^b_i(x)
 & =&  2 \sum_{j\neq i} \im \tr \Bigl\{ T^b \Bigl[  
 U_j(x) U_i(x+a\hat{j}) U^\dagger_j(x+a\hat{i}) U^\dagger_i(x)
 \nonumber \\ && \hspace{2.3cm}
 +\; U^\dagger_j(x-a\hat{j}) U_i(x-a\hat{j}) U_j(x+a\hat{i}-a\hat{j})
    U^\dagger_i(x)  
 \Bigr]\Bigr\} \,.
\label{eq:force}
\end{eqnarray}
What is important here is that the spatial link
variables are compact.  Therefore the size of
the time derivative of ${\cal E}_i^b$ is bounded.  
At sufficiently small $\betaL$ (strong coupling) 
the link matrices become essentially random elements of the
group, and the typical size of $|a \partial_t {\cal E}^b_i|$ saturates.  
On the other hand, the mean-squared value, 
$\langle|{\cal E}_i^b|^2\rangle \sim 1/\betaL$, does not saturate but
increases linearly as $1/\betaL$ is made large.

In this regime, the electric fields feel an essentially random force of fixed
mean-squared value, and evolve much like heavy particle velocities in 
classical Langevin dynamics.  
The time scale for the link matrices $U_i$ to
rotate by an ${\cal O}(1)$ angle is 
$t/a \sim 1/|{\cal E}_i^b| \sim \betaL^{1/2}$.
The force on the electric field, \Eq{eq:force}, involves a product of 
four links which each rotate independently at (generically) irrationally
related frequencies; the product of four such randomly rotating group 
elements should show no periodicity or quasi-periodicity.  Therefore
the coherence time of the random force on ${\cal E}_i^b$ 
is set by the time for a link
matrix to rotate by an ${\cal O}(1)$ angle, i.e.\ just 
$t/a \sim \betaL^{1/2}$.
A random variable ${\cal E}_i^b$ 
with mean squared value $|{\cal E}_i^b|^2 \sim \betaL^{-1}$,
experiencing a random force of magnitude $\sim 1$ with a coherence time
$\sim \betaL^{1/2}$ behaves as 
$\langle {\cal E}_i^b(t) {\cal E}_i^b(0) \rangle \sim 
\langle|{\cal E}_i^b|^2\rangle\exp(-|t|/\tau)$ 
where $\tau \sim \betaL^{-3/2}$.
Integrating over $t$ and inserting
$\langle|{\cal E}_i^b|^2\rangle \sim \betaL^{-1}$, we conclude that
$\kappa \sim \betaL^{-5/2}$ for $\betaL \ll 1$. 
%
%
This description also predicts that the support of
$\kappa_\rmi{latt}(\omega)$ should become narrow with width 
$a\omega \sim a/\tau \sim \betaL^{3/2}$.

However we emphasize that this behavior 
is an artifact of the electric fields being
non-compact while the gauge links $U_i$ are compact. 
Such a disparity is absent in the quantum theory so the effect is
an artifact of the classical lattice discretization.  
Concretely, we find a different qualitative behavior 
in the small-$\betaL$ regime of the ``improved''
description (cf.\ \fig\ref{fig_imp}). Therefore we believe
that the behavior of $\kappa_\rmi{latt}$ 
in the small-$\betaL$ regime has nothing 
to do with real QCD.


\end{document}